# Putting Things in Context: Securing Industrial Authentication with Context Information[1]

**Simon Duque Anton[1], Daniel Fraunholz[1], Christoph Lipps[1], Khurshid Alam[1], and Hans Dieter Schotten[1, 2]**

*1: German Research Center for Artificial Intelligence, Germany*
*2: University of Kaiserslautern, Germany*

## ABSTRACT

The development in the area of wireless communication, mobile and embedded computing leads to significant changes in the application of devices. Over the last years, embedded devices were brought into the consumer area creating the Internet of Things. Furthermore, industrial applications increasingly rely on communication through trust boundaries. Networking is cheap and easily applicable while providing the possibility to make everyday life more easy and comfortable and industry more efficient and less time-consuming. One of the crucial parts of this interconnected world is sound and secure authentication of entities. Only entities with valid authorisation should be enabled to act on a resource according to an access control scheme. An overview of challenges and practices of authentication is provided in this work, with a special focus on context information as part of security solutions. It can be used for authentication and security solutions in industrial applications. Additional information about events in networks can aid intrusion detection, especially in combination with security information and event management systems. Finally, an authentication and access control approach, based on context information and -

---

depending on the scenario - multiple factors is presented. The combination of multiple factors with context information makes it secure and at the same time case adaptive, so that the effort always matches, but never exceeds, the security demand. This is a common issue of standard cyber security, entities having to obey strict, inflexible and unhandy policies. This approach has been implemented exemplary based on RADIUS. Different scenarios were considered, showing that this approach is capable of providing flexible and scalable security for authentication processes.

# 1    Introduction

Interconnectivity, characterised by ad-hoc access to computing systems, is a property of the Industrial Internet of Things (IIoT). Network- and cloud-based resources, in addition to mobile and handheld computing, motivate a paradigm shift in authentication and access control. In the consumer area, users want to access the same resource from different devices in different places with connections that break trust boundaries. Furthermore, access rights should be transferred between different devices. In the industrial area, users want to access different production entities and networks without re-authentication. Access rights should be traversed as well, from one accessed device in a certain network to another. These requirements are easy to fulfill with standard means. However, with the increase in interconnected devices, network-based crime has drastically increased as well. Several recent and not-so-recent botnets aimed at IoT-devices, like the infamous *Mirai*-botnet (Antonkakis et al. 2017), or its successor, the *Satori*-botnet (Paganini 2018). In addition to that, the industrial landscape is becoming more and more of a target for cyber attacks (Duque Anton et al. 2017a). Cyber crime has a greater revenue than drug trafficking since 2009 (Dethlefs 2015; Symantec 2009). This trend creates demand for strong authentication and access control, as it is the foundation of any further interaction and needs to be fulfilled in order to meet any other security objectives (Schaefer 2016). At the same time, typing a strong user name and password-combination on a cell phone, for example, is bothersome and inflexible. Authentication should be easy to use and at the same time secure. Additionally, connectivity through trust boundaries and heterogeneous networks demand for novel kinds of intrusion detection. Many industrial protocols do not contain means for encryption or authentication by default.

The contribution of this work consists of two parts: First, providing sound and secure authentication, based on information about circumstances, so-called situational awareness is addressed in this work in Section 3. Current research areas, challenges and solutions for authentication are evaluated in Section 5. Second, an exemplary implementation of a context-based authentication system is implemented and evaluated, derived of a use case that fits both classic home and office networks, as well as industrial applications.

The remainder of this work is structured as follows: A short overview of existing authentication methods is provided in Section 2. Section 3 provides a concept for integration of situational awareness into authentication. The proposed approach is presented in Section 4 and evaluated with three scenarios. In Section 5, an overview of current research in the field of multi-factor and context-based authentication and authorization is given. Future research directions are presented in Section 6, this work is concluded in Section 7.

# 2    Authentication in a Nutshell

The process of authentication can be divided into three steps:
- Identification,
- Authentication and
- Authorisation.

In the first step, identification, an entity makes an assertion about who it claims to be. In the next step, authentication[2], this assertion is evaluated. Lastly, after establishing the identity, access rights are granted accordingly in the authorisation phase. These three steps are summarised in this section. First, identification is discussed in Subsection 3.1. There are several well-established and commonly used authentication mechanisms. According to Turner, they can be divided into three different kinds of authentication factors, as described in Subsection 3.2 and four different types of authentication procedures (Turner 2016), as presented in Subsection 3.3. The authentication mechanisms presented here are used for authentication of human users. In contrast to human users, machines and services can be users that need authentication in IIoT scenarios as well. Finally, authorisation is discussed in Subsection 3.5

## 2.1 Identification Process

The identification of an entity is the first step of the authentication process. Only if an authentication provider knows the identity of a given entity, the according access rights can be assigned. Access rights of a given entity must only be granted to this entity (Schneier 2000). In order to do so, an identity is first requested. The entity makes a claim about its identity. In order to evaluate this claim, the entity has to be known to the authentication provider beforehand. Otherwise, it must register. Methods to make a claim about an identity are usernames, but can also be extended to barcodes, Radio Frequency Identification (RFID) token or Hardware Security Modules (HSMs) and Trusted Platform Modules (TPMs) that are integral to smartcards (International Organization for Standardization 2011). Furthermore, there are use cases where it is necessary to authenticate entities that have not been previously known to the authentication provider, called *Plug & Trust* (Ambekar et al. 2012) or *Trust On First Use (TOFU)* (Gultsch 2016). The identification and authentication methods for such use cases are based on properties that can be verified without prior knowledge of the specific entity.

---

[2] Authentication is used ambiguously in this work, describing the process of proving an identity as well as the whole process

## 2.2    Authentication Factors

The different authentication factors can be grouped as follows:
- Knowledge factors,
- Possession factors and
- Inherence factors.

This differentiation is used to categorise authentication factors of human entities. In the context of IIoT, machines often need to authenticate to services or other machines. The differentiation between these factors, however, is different than that of humans. Concepts, such as knowledge and possession, do not work for machines.

*Knowledge factors* are most commonly used in form of passwords or Personal Identification Numbers (PINs). They prove the identity of a user by providing knowledge that, by default, only the given user should have. They are usually easy to change in case they are compromised. Given the handling is done in a secure manner by the service provider requesting the passwords, they are relatively secure. Good handling includes storage as salted hashes only and encrypted transmission. On the downside, remembering passwords is tedious, especially if a user has to remember many of them. So-called key-managers for mobile- and desktop-systems, such as *LastPass (LogMeIn 2018)*, increase the usability.

*Possession factors* are used to prove the identity of a user with a certain item only the user should have. Common examples are *RSA SecureID* (RSA Security LLC 2018) or ID cards, for example by *Siemens* (2013). One Time Passwords (OTPs) (Haller et al. 1998) can be considered a combination of knowledge and possession factor. They contain a secret that cannot be extracted by an attacker and are usually very secure. Furthermore, their loss is easily detected, so a compromise of security can be detected in a timely manner. In contrast to that, if a password has been cracked, the user has no way of knowing.

*Inherence factors* are usually associated with biometric information, such as finger prints or retina scans. This kind of factor is meant to be used to unambiguously identify a user. It is a characteristic that cannot be changed. The most prevalent disadvantage lies in the fact that, once compromised, this factor can never be used again, as it is fixed for the lifetime of a user. That means, once the biometric factor of a person is publicly available, forgeries can be made with sufficient effort. Prominent examples are the forgery of German chancellor *Dr. Angela Merkel's* retina scan and German Secretary of Defense, *Dr. Ursula von der Leyen's* finger prints (Krempl 2014). Both replicas were made solely from pictures that were published in press releases. This impressively shows the dangers of inherence factors.

## 2.3    Types of Authentication

According to *Turner*, four different types of authentication can be distinguished (Turner 2016). In contrast to the authentication factors describing how authentication is performed, authentication types describe the combination of different factors. They are listed below in order of increasing security:

 – Single-factor authentication
 – Two-factor authentication
 – Multi-factor authentication
 – Strong authentication

*Single-factor authentication* is based on one authentication factor, e.g. a password or PIN. If this factor is flawed, the approach does not provide security. *Two-factor authentication* takes two authentication factors in order to determine the identity of a user. It is harder to break, as two factors need to be compromised. However, if the required input is based on the same factor, e.g. a possession factor like an Identifier (ID) badge, this approach is only as secure as the given factor. *Multi-factor authentication* is based on a multitude of factors, but at least three. As with two-factor authentication, if the input is based on the same factor, the mechanism is only as secure as the given factor. *Strong authentication* is two- or multi-factor authentication, but with so-called strong factors. Strong factors must not depend on other factors used by the authentication process. Furthermore, they must be designed in a fashion that they cannot be generated by an attacker with knowledge of the authentication process.

## 2.4    Context as a Factor

Context information can be considered as an authentication factor. It provides information about the status of an entity, therefore it can be used as a factor or increase the measure of trust (Bhatti et al. 2004; Hulsebosch et al. 2005; Lenzini 2009). Location information can be used to determine whether a user is within a perimeter. If not it is less likely that the user is affiliated with the company owning the perimeter. In the same way time can be a factor. Enterprises have common working hours during which access requests to resources are common. If a request occurs outside those hours, the credibility of the requesting entity is lower. Another application of context is access control based on necessity and emergency. One possible application of access control is granting entry to locked doors for emergency systems only if an alarm was triggered. *Choi et al.* consider context for risk-based access to sensitive medical information (Choi et al. 2015).

## 2.5    Authorisation Schemes

Authorisation and access control describe the process of granting rights to entities, once their identities have been confirmed. Depending on the access control policies, different users or types of users have certain permissions in a network. Access control and authorisation inherently only works with secure and strong authentication. If there is no possibility to reliably determine the identity of a user, access control and authentication are of no value. According to *Musa*, there are four general types of access control (Musa 2014):

- Mandatory Access Control (MAC)
- Discretionary Access Control (DAC)
- Role-Based Access Control (RBAC)
- Rule-Based Access Control

*MAC* grants access rights and permissions based on strict schemes that arise from user level and the classification of data. *DAC* allows the owner of the data to specifically grant rights for each individual user. *RBAC* distributes access rights and permissions based on the role of the user. Each user is assigned one role, which has different rights for different types of data and different permissions for several actions. *Rule-Based Access Control* allows granting permissions based on some rules that can also relate to environment and context information.

# 3 Concept for Context Integration

The main changes brought to industry in the course of the IIoT are based on an increase of connectivity and the distribution of computations. In the beginning of Supervisory Control And Data Acquisition (SCADA) systems, industrial networks were physically separated from public networks access to devices had to be obtained either locally or by using dedicated control terminals. Now, operators often switch between machines and tasks often, sometimes access is requested from remote locations. Due to the availability of resources over public networks, strong authentication is required. On the other hand, strong authentication is often tedious to use for human operators. Common authentication mechanisms are presented in Section 2. The concept proposed in this work is based on two aspects. First, the strength of authentication is dependent on the action that is requested. E.g. if an operator aims to perform security critical tasks, such as reconfiguration of a device, a stronger authentication is required than if the operator only requests to read status information. Second, situational awareness is taken into account to evaluate the risk of a request. In practice, local access can be deemed safer than remote access, as local access requires physical presence that is enforced by physical security. In addition to that, time of the request provides information of its soundness. Requests during working hours are very probable while requests during off time are less likely, but still possible. In taking this into account, the strength of authentication can be adapted dynamically. The scheme is pictured in Figure 1. Context variables describe the circumstance of a request, providing situational awareness. The authentication methods are inherent to a requester. A validator decides upon the strength of authentication required based on the situation and the requested action or resource. An exemplary implementation of this system is presented in the subsequent section.

# 4 A Dynamic Model

In this section, the use case is described in the Subsection 4.1. The architecture is presented in Subsection 4.2, followed by the evaluation in Subsection 4.3 and a discussion in Subsection 4.4.

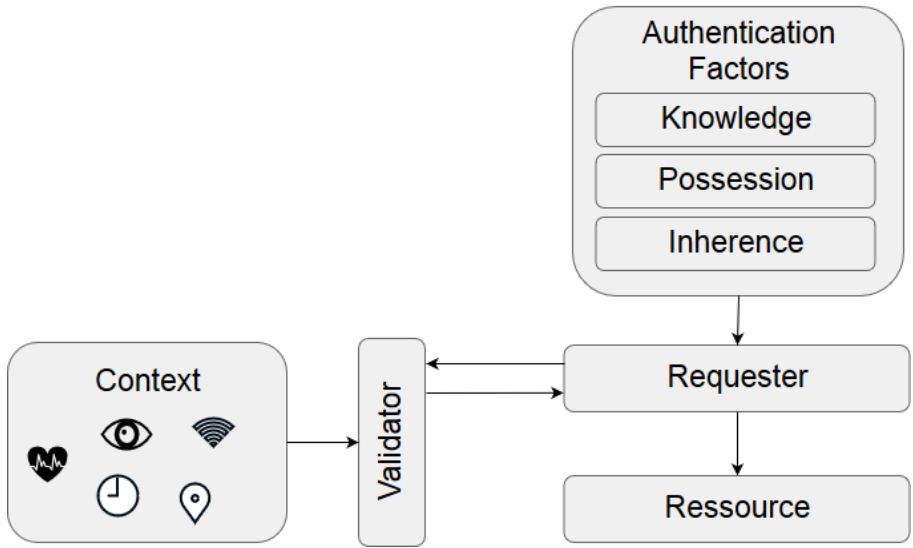

FIGURE 1: AUTHENTICATION CONCEPT

## 4.1 Use Case

The use case evaluated in this work is exemplary for existing requirements in industrial and office network requirements. It focusses on user access to resources under different circumstances with requests to perform a set of actions. The action can either be a request for default or for root access to a resource. The situational awareness is derived from information about time and location, where location is obtained from the IP address. This means a user can either be inside a trusted network, or not. The time can either be during common working hours or not. If a user requests any action, default or root, outside the trusted network or outside common working hours, high security is required. High security means two factors in the course of the implementation presented in this work. If a user requests default access during normal working hours and from inside a trusted network, the situation is deemed secure so that low security is sufficient. Low security means only one authentication method is required. If a user requests root access to the resource, high security is required as this kind of access motivates a high level of security.

The combinations of circumstances, requested actions and resulting authentication factors is listed in Table 1. The access control is a Rule-Based approach, as described in Section 3.2.

TABLE 1: REQUIRED SECURITY LEVEL FOR COMBINATIONS OF CONTEXT AND ACTION

| Context | Request | Low Sec. | High Sec. |
|---------|---------|----------|-----------|
| Normal working hours and on site | default access | ✔ | ✘ |
| Normal working hours and on site | root access | ✘ | ✔ |
| Outside working hours or off site | default access | ✘ | ✔ |
| Outside working hours or off site | root access | ✘ | ✔ |

## 4.2 Architecture

The proposed authentication scheme is described by a flow diagram shown in Figure 2. First, the system checks if the user is already authenticated either with default or with root access. If the user is neither, one factor is requested by the system. For our proof of concept implementation, we decided on a username-password combination that needs to be provided by the user. If the provided credentials are incorrect, the authentication request is denied. Else, the context provided is checked. For the implementation presented in this work, location, based on IP and time, based on the authentication server time, are employed. If those context variables are during common working hours, or the location is outside the office area, a second factor, implemented as an OTP, is requested. If the context is plausible, but the user is requesting root permissions, an OTP is requested as a second factor as well. Furthermore, it is possible for a user with default access rights to request root permission after some time, if need be. The algorithm then requests the second factor only at that time and allows for an update of access rights.

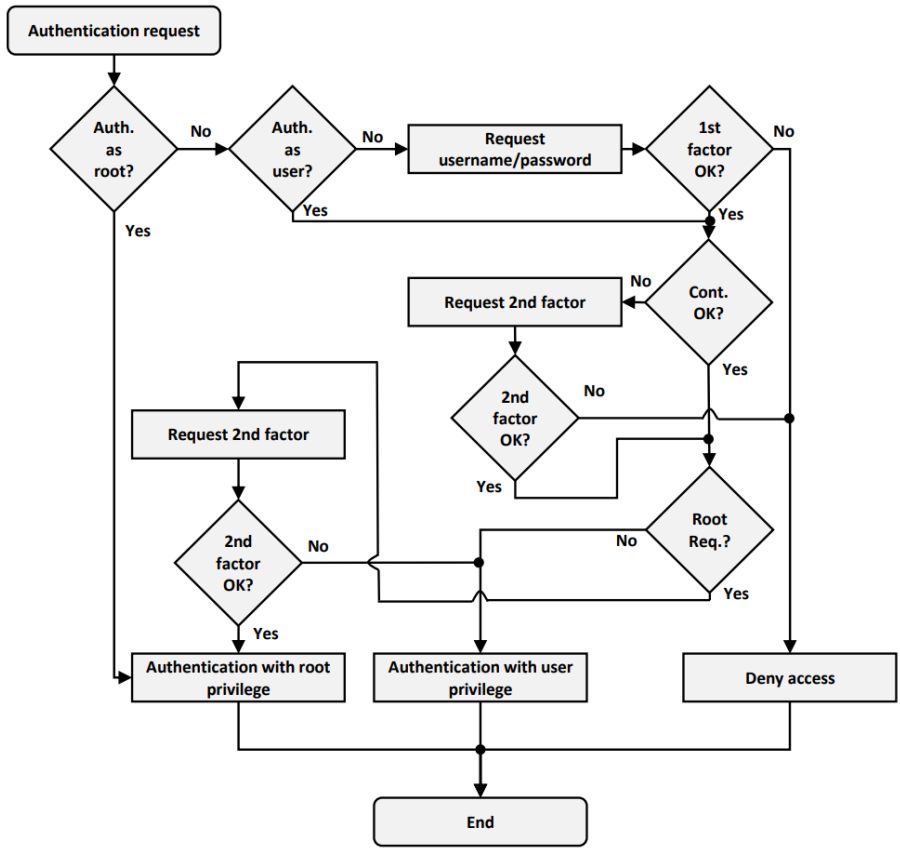

FIGURE 2: AUTHENTICATION MODEL ARCHITECTURE

## 4.3    Evaluation

In this section, the exemplary implementation of the considered scenarios is presented. First, the used software components are explained in Subsection 4.3.1. After that, the behaviour of the presented implementation when faced with different conditions and scenarios is described in Subsection 4.3.2.

### 4.3.1  Implementation

The exemplary implementation has been done based on Remote Authentication Dial-In User Service (RADIUS) (Rigney et al. 2000). RADIUS is an open-source protocol used for authentication users in networks. There are implementations publicly available, making it easy to adapt. The libraries for user name-password-combinations are publicly available. For user name and password combination, the Password Authentication Protocol (PAP)-library was used. In case an OTP was needed, the Short Message Service OTP (SMSOTP) module was used. The structure of the authentication engine can be found in Figure 3.

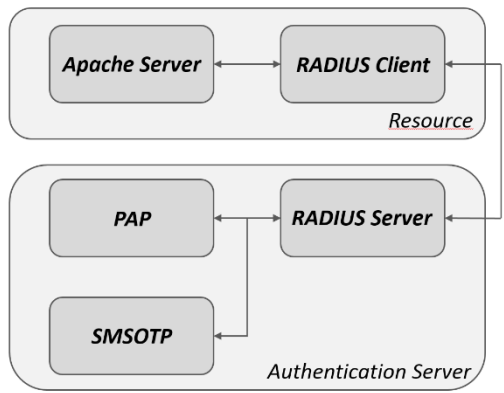

FIGURE 3: STRUCTURE OF THE AUTHENTICATION ENGINE

RADIUS provides encryption for the transmission of passwords using PAP and SMSOTP. The algorithm, described in depth by *Hill* (2001), is based on a shared secret *S*. It is summarised in the following equation:

$$c_1 = p_1 \oplus MD5(S + RA)$$
$$c_2 = p_2 \oplus MD5(S + c_1)$$

$$\dots$$
$$c_n = p_n \oplus MD5(S + c_{n-1})$$

The cypher text blocks $c_x$ for x $\in$ {1, ..., n} are concatenated and transmitted to the RADIUS server as the password-attribute. Cypher text block $c_i$ depends on cypher text block $c_{i-1}$, the first block $c_1$ is initialised with a Request Authenticator *RA*, as described in the first line of Equation 1. The $p_x$ for x $\in$ {1, ..., n} are 16-octet blocks of the user password. *MD5* is a hashing algorithm. The presented RADIUS-modules have been extended with modules that get the context on the server-side in order to determine the required security level. Furthermore, a module has been written that re-requests authentication if high security is demanded. This request demands a OTP. The resource to be authenticated to is an *apache* web-server (Apache Foundation 2018) that hosts documents. Default and root permissions can be granted to a user.

### 4.3.2 Scenarios
In order to evaluate the implementation, three different scenarios were considered. In the first scenario, low security access is considered. The authentication framework only requests user name and password. In the second and third scenario, high security is considered. The second scenario describes a user requesting root access to the resource, while the third scenario describes a user requesting default access, but with context information that requires an additional authentication factor. The information flow for standard and root authentication can be found in Figure 3. As the procedure remains the same, whether the user demands root access right away or demands it after some time, no specific scenario where the access rights are increased after some time is considered.

*First scenario:* The user is prompted with a request for user name and password. After providing these, default access to the resource is granted. This flow can be seen in Figure 4 by omitting the part marked as "root access".

*Second scenario:* The user is prompted with a request for user name and password. After providing these, the system figures that root access was requested that demands for high security. Therefore, a second factor is requested. After providing it, the user is granted root access to the resource. This is described as the full flow in Figure 4.

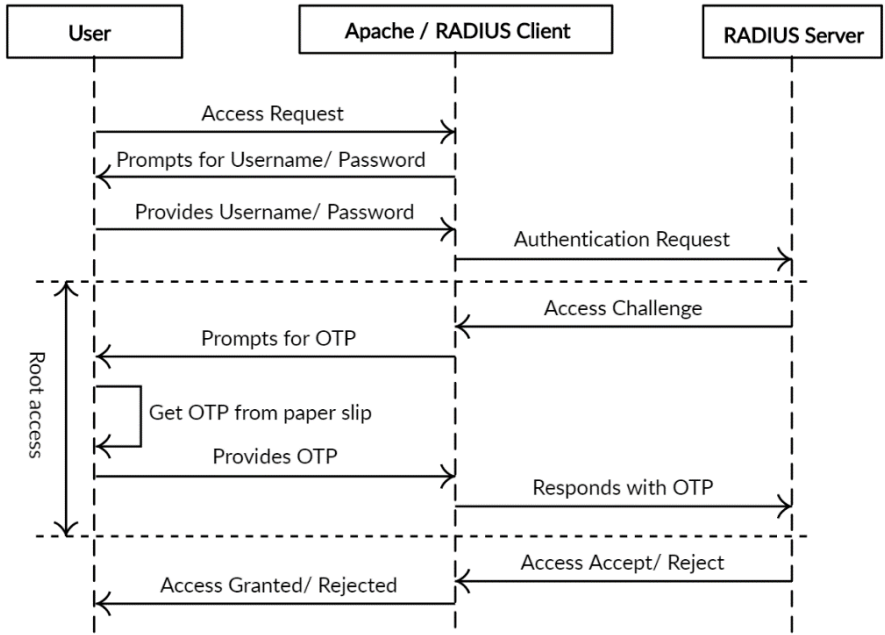

FIGURE 4: AUTHENTICATION SEQUENCE

*Third scenario:* The user is prompted with a request for user name and password. After providing these, the system checks the context variables and determines that high security is required. A second factor is requested. After providing it, the user is granted default access to the resource. This, as well, is described as the full flow in Figure 4.

## 4.4 Discussion

In this section, a model for context-based was introduced and evaluated. The extend is relatively simple, containing only two authorisation levels, namely standard and root, as well as two context factors, namely IP-based location and time. The implementation of this model shows the feasibility of the concept. There are advantages to the proposed solution, as well as challenges that have not yet been addressed. First, RADIUS-based solutions are easy to implement and integrate, as it is a well-established tool for authentication with open source code, providing the possibility for extensions. Second, the context variables that have been employed are easily collected, keeping the additional effort small. No further sensors are needed to gather all information, as they are readily available.

On the other hand, there are more novel solutions for authentication and authorisation available, addressing challenges that did not play a major role in the development of RADIUS, such as roaming in WLAN networks and authentication in untrusted networks. Furthermore, the context information can be spoofed with medium effort. IPSec and protocols can be used to secure the location factor. Additional information, such as hardware security tokens provide more flexibility and a high level of security as a second factor. However, they require additional hardware and implementation effort. Finally, the context variables that have been used are limited, there is an abundance of possible context that can be integrated into authentication solutions. Possible context information is discussed in Section 5.

From the authorisation point of view, a scenario with two roles, default and admin, is common for office, and sometimes home, use. However, to meet requirements of current industrial requirements, more roles are required. Furthermore, timing constraints and time-based access is important for actions that are scheduled in a predefined time. Categories of context factors, examples and factors that have been used in this work are listed in Table 2.

TABLE 2: CONTEXT FACTORS

| Context Category | Example | Used in this work |
|---|---|---|
| **Time** | NTP-based, server-based, host-based | Server-based |
| **Location** | GPS, IP, Proximity, Wireless (WLAN, Bluetooth, …) | IP |
| **Event** | Emergency, Condition | - |
| **Entity-based** | Histroy, Chain of Trust, Biometric/physical properties | - |

# 5    Related Work

In this section, related work in authentication is presented. Apart from the well-established classic authentication mechanisms as described in Section 3, there is ongoing research in the field of sound and secure authentication. Several works address the aspect of trust and computation of a trust level in authentication and access control. *Bhatti et al.* present a trust-based context-aware access control model for web-services that address shortcomings of classic approaches (Bhatti et al. 2004). *Lima et al.* compute trust, based on recommendations in mobile and pervasive environments to provide algorithms that meet demands in dynamic and security for mobile computation (Lima et al. 2011). *Rene and Gellersen* present a method to pair devices based on shaking them (Mayrhofer and Gellersen 2007). Furthermore, *Mayrhofer* worked on context-based authentication for spontaneous and ubiquitous devices (Mayrhofer 2006). In addition to single-factor authentication schemes, multi-factor authentication is a widely regarded topic in research, e.g. the work of *Dasgputa et al.* (2017). They provide an exhaustive overview over authentication processes and an extensive insight into the advances in this field. There are multi-factor authentication schemes with certain policies, like the privacy-preserving biometric authentication as presented by *Bhargav-Spantzel et al.* (2017). Many multi-factor authentication mechanisms address the issue of mobile and distributed devices and the resulting necessity of secure authentication. *Glynos et al.* present an approach for Mobile Ad Hoc Networks (MANETs) to prevent spoofing and impersonation attacks (Glynos et al. 2005). The importance of trust in MANETs as well as the difficulties in computing it is described in the work of *Govindan and Mohaparta* (2012). *Huang et al.* work on multi-factor authentication for fragile communications, e.g. when authentication servers are often unavailable (Huang et al. 2014). They address the problem of slow or unavailable remote authentication servers and present an authentication protocol that speeds the authentication process with a remote entity up. The specific dangers of cloud computing are considered by *Banyal et al.* (2013). A framework consisting of authentication mechanisms that are compatible with established schemes and methods for access management is proposed. *Aloul et al.* present an approach that increased the level of authentication by using mobile phones, for applications such as banking (Aloul et al. 2009). Furthermore, as authentication and access control is employed in a variety of fields, from online banking to accessing one's cell phone, some authentication processes are more critical than others. *Jae-Jung and Seng-Phil* present a method of assessing the risk and trustworthiness of authentication methods and provide insight as to when they provide a sufficient amount of security (Jae-Jung and Seng-Phil 2011).

In addition to that, context-based authentication and access control is a widely regarded topic. It is addressed by *Hulsebosch et al.* (2005). They investigate the possibilities of access control based only on situational information. Furthermore, they explore the possibilities of providing anonymous services due to the obtained information. *Kindberg and Zhang* look into context authentication over constrained channels (Kindberg and Zhang 2001). They propose an algorithm to determine the position of a user based on physical channel information. *Kinberg et al.* present a model that derives context information about the location based on characteristics of the communication channel (Kindberg et al. 2002). Proximity awareness as a means of context is considered by *Lenzini* (2009) and *Bardram et al.* (2003). *Bardram et al.* focus explicitly on pervasive computing as a modern day challenge. They propose context awareness to determine a user's location in order to grant or deny access. *Primo et al.* propose a framework for context-aware authentication based on the position a user holds the smartphone by extracting position information of the accelerometer (Primo et al. 2014). *Lenzini* introduces a system that is aware of users in its proximity and able to identify and authenticate these users. Furthermore, a level of trust that the user is actually at the assumed position is calculated. *Hayashi et al.* propose a solution similar to the one implemented in this work. They use context information about the user, such as location, in order to determine security factors and select the authentication scheme accordingly (Hayashi et al. 2013). A privacy-enhanced access control scheme that optionally integrates context in pervasive environments is developed by *Ren and Lou* (2007). *Hohlbein and Teufel* address the issue of authorised persons misusing the system by introducing a role-based access control scheme that uses a need-to-know basis (Holbein and Teufel 1995). In addition to that, context information can be used to determine the security level of a network, as proposed by *Duque Anton et al.* (2017b), or to perform attribution of attacker types, as presented by *Fraunholz et al*. (Fraunholz et al. 2017a; Fraunholz et al. 2017c).

The challenges faced by authentication and authorisation, as well as solutions proposed by works presented above are listed in Table 3.

TABLE 3: CHALLENGES IN AUTHENTICATION AND AUTHORISATION

| Challenges | Addressed |
|---|---|
| Trust computation | (Bhatti et al. 2004; Lima et al. 2011; Govindan and Mohapatra 2012) |
| Authentication for critical services | (Aloul et al. 2009) |
| Multi-factor authentication for MANETs | (Glynos et al. 2005; Govindan and Mohapatra 2012) |
| Privacy-preserving multi-factor authentication | (Bhargav-Spantzel et al. 2017) |
| Multi-factor authentication for pervasive environments | (Ren and Lou 2007; Mayrhofer 2006; Mayrhofer and Gellersen 2007; Banyal et al. 2013) |
| Multi-factor authentication for fragile communication | (Huang et al. 2014) |

| Risk assessment of authentication methods | (Jae-Jung and Seng-Phil 2011) |
|---|---|
| Context-aware authentication and authorisation | (Holbein and Teufel 1995; Bardram et al. 2003; Kindberg et al. 2002; Hulsebosch et al. 2005; Mayrhofer and Gellersen 2007; Ren and Lou 2007; Lenzini 2009; Hayashi et al. 2013; Primo et al. 2014) |
| Authorisation over constrained channels | (Kindberg and Zhang 2001; Kindberg et al. 2002) |
| Survey | (Dasgupta et al. 2017) |
| Context-aided intrusion detection | (Fraunholz et al. 2017a; Fraunholz et al. 2017b; Fraunholz et al. 2017c; Duque Anton et al. 2017b) |

# 6 Future Research Directions

As a next step towards the integration of situational awareness into authentication solutions, different types of authentication factors will be evaluated. Especially interesting are factors that provide strong authentication, while needing no user interaction whatsoever. A worker in a factory, for example, that carries a tablet PC in order to wirelessly control devices, could be identified by a RFID-tag around his neck. If strong security is needed, the tabled PC could automatically scan his iris to get a second factor. Furthermore, physical attributes of devices or wireless channels can be used for strong, automated authentication and encryption (Lipps et al. 2018; Weinand et al. 2017). Furthermore, Integrating context into an aggregation model can aid in intrusion detection, especially for heterogeneous networks (Duque Anton et al. 2017c).

# 7 Conclusion

The presented context-based multi-factor authentication approach provides an additional level of security in comparison to standard authentication mechanisms as discussed in Section 3. For the exemplary implementation well-established authentication factors have been used. They could easily be extended or replaced in order to obtain on-demand security with strong authentication. The crucial point of the proposed framework is the flexibility and adaptability to different security standards. For less security critical tasks, one factor authentication is used, while more critical tasks require two factors. In the proposed framework, context information is considered in order to decide whether a task is critical or not, as the criticality also depends on the circumstances of the task. Especially the dynamic increase of access rights after providing a second factor makes this approach well suited for a dynamic environment. The principle of least needed access rights applies, so that the second factor is only necessary if the policy demands it. Several existing technologies were combined in order to keep the system easy-to-use, while still capable to adapt to the environment.

**Acknowledgements** This work has been supported by the Federal Ministry of Education and Research of the Federal Republic of Germany (Foerderkennzeichen KIS4ITS0001, IUNO). The authors alone are responsible for the content of the paper.

## KEY TERMS

- Context: Information about the conditions and current status of an environment
- Identification: The claim an entity makes about its identity
- Authentication: The process of proving an identity towards a system
- Multi-factor Authentication: The process of using different mechanisms at once to prove identity towards a system with a higher security
- Strong Authentication: The process of proving the identity towards a system by using independent authentication mechanisms
- Authorisation: The process of providing access rights to a resource by an authority based on a set of rules.
- Industrial Internet of Things: A concept that builds on interconnectivity and distributed computation to increase flexibility and productivity of industrial applications and to reduce cost and maintenance

## BIOGRAPHICAL NOTES

**Simon Duque Anton** obtained his Diploma in Information Technologies in 2015 at the University of Kaiserslautern. Born in Hamburg, Germany in 1989, he started working at the German Research Center for Artificial Intelligence in Kaiserslautern after that, where he pursues a Ph.D. in the field of machine learning and artificial intelligence. His research covers the application of machine learning methods on industrial intrusion detection and network security.

**Daniel Fraunholz** is a Researcher and Ph.D. candidate at the Intelligent Networks Research Group at the German Research Center for Artificial Intelligence in Kaiserslautern since 2015. Born in Stuttgart, Germany in 1992, he received his Bachelor and Master degree from Heilbronn University of Applied Sciences in Electronic Systems Engineering in 2014, respectively 2015. His major research interests are network security, intrusion detection and deception systems.

**Christoph Lipps** graduated in Electrical and Computer Engineering at the University of Kaiserslautern. He received his Master degree in Communication Systems in 2016. Born in Pirmasens, Germany in 1986, he started working as a Researcher and Ph.D. candidate at the German Research Center for Artificial Intelligence in Kaiserslautern. His research focuses on Physical Layer Security, Physically Unclonable Functions and entity authentication.

**Khurshid Alam** obtained his Master's degree in Applied Computer Science in 2018 at the University of Kaiserslautern and started working at Intelligent Networks Research Group of the German Research Center for Artificial Intelligence since then. His research area covers industrial wireless communication and network security.

**Hans Dieter Schotten** is a full professor and head of the chair for Wireless Communications and Navigation at the University of Kaiserslautern. In addition to that, he is Scientific Director of the Intelligent Networks Research Group of the German Research Center for Artificial Intelligence (DFKI). He received his Ph.D. in Electrical Engineering from the RWTH Aachen in 1997. He was a research group leader there before changing into industry. At Ericsson, he held the position of Senior Researcher, after that he held the position of Director of Technical Standards at Qualcomm. His topics of interest are wireless communication and 5G.